\documentclass{iaus}
\usepackage{graphicx}

\newcommand{\mic}{\hbox{$\mu$m}}

\newcommand{\mdust}{\hbox{$M_{\mathrm{d}}$}}

\newcommand{\sfr}{\hbox{$\psi$}}
\newcommand{\ssfr}{\hbox{$\psi_{\mathrm S}$}}

\newcommand{\iras}{\hbox{\it IRAS}}
\newcommand{\iso}{\hbox{\it ISO}}
\newcommand{\galex}{\hbox{\it GALEX}}

\newcommand{\mstar}{\hbox{$M_{*}$}}

\title[A simple model to interpret the SEDs of galaxies]{A simple model to interpret the ultraviolet, optical and infrared SEDs of galaxies}

\author[E. da Cunha, S. Charlot \& D. Elbaz]   
{Elisabete da Cunha$^{1,2}$, St\'ephane Charlot$^{3,4}$, \& David Elbaz$^5$}

\affiliation{$^1$Department of Physics, University of Crete, 71003 Heraklion, Greece \\ [\affilskip]
$^2$ IESL/Foundation for Research and Technology-Hellas, 71110 Heraklion, Greece \\ [\affilskip]
email: {\tt dacunha@physics.uoc.gr} \\[\affilskip]
$^3$ UPMC Univ. Paris 6, UMR7095, Institut d'Astrophysique de Paris, 75014 Paris, France\\ [\affilskip]
$^4$ CNRS, UMR7095, Institut d'Astrophysique de Paris, 75014 Paris, France\\ [\affilskip]
$^5$ Laboratoire AIM, CEA/DSM-CNRS-Universit\'e Paris Diderot, IRFU/Service d'Astrophysique, CEA-Saclay, 91191 Gif-sur-Yvette Cedex, France}

\pubyear{2009}
\volume{262}  
\jname{Stellar Populations -- Planning for the Next Decade}
\editors{G. Bruzual, \& S. Charlot, eds.}
\begin{document}

\maketitle

\begin{abstract}
We present a simple, largely empirical but physically motivated model,
which is designed to interpret consistently multi-wavelength 
observations from large samples of galaxies in terms of physical parameters, such as 
star formation rate, stellar mass and dust content. Our model is both simple and versatile
enough to allow the derivation of statistical constraints on the star formation 
histories and dust contents of large samples of galaxies using a wide range of ultraviolet,
optical and infrared observations. We illustrate this by deriving median-likelihood 
estimates of a set of physical parameters describing the stellar and dust contents of
local star-forming galaxies from the {\it Spitzer} Infrared Nearby Galaxy Sample (SINGS)
and from a newly-matched sample of SDSS galaxies observed
with \galex, 2MASS, and \iras. The model reproduces well the 
observed spectral energy distributions of these galaxies across the entire wavelength 
range from the far-ultraviolet to the far-infrared. We find important correlations between
the physical parameters of galaxies which are useful to investigate the
star formation activity and dust properties of galaxies.
Our model can be straightforwardly applied to interpret observed ultraviolet-to-infrared
spectral energy distributions (SEDs) from any galaxy sample.

\keywords{dust, extinction; galaxies: stellar content; galaxies: ISM; galaxies: statistics.}
\end{abstract}

\firstsection 
\section{Introduction}
             
Combined ultraviolet, optical and infrared data are now becoming available for 
large samples of galaxies. To extract constraints on the stellar populations and ISM of 
galaxies from these multi-wavelength observations requires the consistent modelling 
of the emission by stars, gas and dust.
A standard approach to this problem is to compute the radiative transfer
of the light emitted by evolving stellar populations through gas and dust in the ISM
(e.g. Silva et~al. 1998, Dopita et~al. 2005). Such sophisticated models
provide valuable insight into the detailed emission of individual galaxies.
However, because of the complexity of
radiative transfer calculations, these models are not optimised to derive statistical
constraints on physical parameters from observations of large samples of galaxies.
We present a simple, largely empirical but physically motivated model which can be used to
interpret the mid- and far-infrared SEDs of galaxies consistently
with the emission at ultraviolet, optical and near-infrared wavelengths.
This model allows us to derive statistical estimates on physical parameters
related to the star formation history and dust content of large samples of galaxies.

\section{Description of the model}

We compute the emission by stars in galaxies
using the latest version of the \cite{Bruzual2003} population synthesis code
(Charlot \& Bruzual, in prep.).
The stellar emission is attenuated using the simple
two-component dust model of \cite{Charlot2000}, which accounts for
the fact that stars are born in dense molecular clouds with
typical lifetimes of $10^7$~yr; at later ages, stars migrate to
the ambient (diffuse) ISM (see equation 2 in da Cunha et~al. 2008).
We use an `effective absorption' curve for each
component, $\hat \tau_\lambda \propto \lambda^{-n}$, where the slope $n$
reflects both the optical properties and the spatial distribution
of the dust. We adopt $n=0.7$ for the ambient ISM and $n=1.3$
for the stellar birth clouds.
This prescription allows us to compute the total energy absorbed by
dust in the birth clouds and in the ambient ISM; this energy is
re-radiated by dust at infrared wavelengths.

We define the total dust luminosity
re-radiated by dust in the birth clouds and in the ambient ISM as
$L_\mathrm{d}^\mathrm{\,BC}$ and $L_\mathrm{d}^\mathrm{\,ISM}$,
respectively. The total luminosity emitted by dust in the galaxy
is then $L_\mathrm{d}^\mathrm{\,tot} = L_\mathrm{d}^\mathrm{\,BC} + L_\mathrm{d}^\mathrm{\,ISM}$.

We distribute $L_\mathrm{d}^\mathrm{\,BC}$ and $L_\mathrm{d}^\mathrm{\,ISM}$
in wavelength over the range from 3 to 1000~\mic\ using four main components:
\begin{itemize}
\item the emission from polycyclic aromatic hydrocarbons (PAHs;
i.e.~mid-infrared emission features), \item the mid-infrared
continuum emission from hot dust with temperatures in the range
130--250~K, \item the emission from warm dust in thermal
equilibrium with adjustable temperature in the range 30--60~K,
\item the emission from cold dust in thermal equilibrium with
adjustable temperature in the range 15--25~K.
\end{itemize}
In stellar birth clouds, the relative contributions to
$L_\mathrm{d}^\mathrm{\,BC}$ by PAHs, the hot mid-infrared 
continuum and warm dust are kept as adjustable parameters.
These clouds are assumed not to contain any cold dust. In the
ambient ISM, the contribution to $L_\mathrm{d}^\mathrm{\,ISM}$
by cold dust is kept as an adjustable parameter. The relative
proportions of the other 3 components are fixed to the 
values reproducing the mid-infrared cirrus emission of the Milky
Way. We find that this minimum number of 
components is required to account for the infrared SEDs of galaxies
in a wide range of star formation histories (see da~Cunha et~al. 2008
for details).

\section{The star formation activity and dust content of local galaxies}

The model summarised in the previous section allows us to derive statistical constraints on the physical
properties of galaxies from ultraviolet, optical and infrared observations.
We build a comprehensive library of stochastic models covering a wide
range of star formation histories, metallicities, dust attenuation, dust temperatures and
fractional contributions by different dust components to the total infrared luminosity.
For each model in the library, we compute the predicted ultraviolet, optical and infrared
fluxes in the \galex, SDSS, 2MASS, {\it Spitzer}, \iras, \iso\ and SCUBA bands. We compare
these predictions with observations of two local samples of galaxies: the {\it Spitzer} Infrared
Nearby Galaxy Sample (SINGS, Kennicutt et~al. 2003), and a sample of 3258 SDSS galaxies
with complementary photometric observations by \galex, 2MASS and \iras\ (da Cunha et~al. 2009).

For each observed galaxy, we assess how each model in the stochastic library (characterised by a set
of randomly drawn physical parameters) fits the observed SED by computing the $\chi^2$ goodness-of-fit.
We build the likelihood distribution of any physical parameter by weighting
the value of that parameter in each model of the library by the probability $\exp(-\chi^2/2)$. Our final estimate
of the parameter is the median of the likelihood distribution and the associated confidence interval is the
16th--84th percentile range.  

\subsection{SINGS sample}

\begin{figure}
\begin{center}
\includegraphics[width=0.9\textwidth]{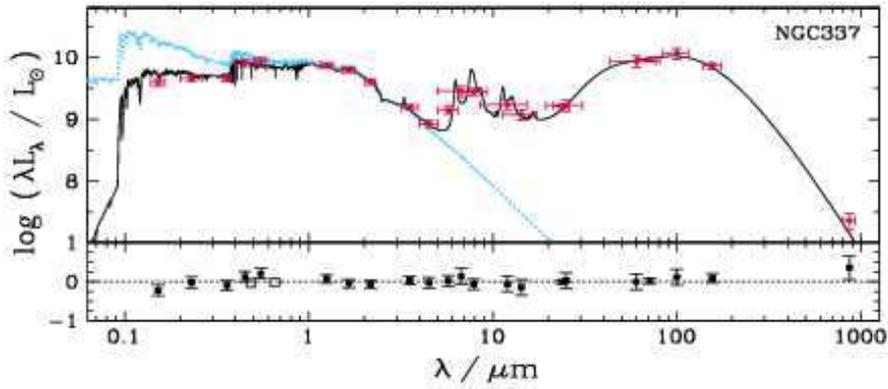} 
\vspace*{-0.2 cm}
 \caption{Best-model fit to the observed SED (red squares) of the SINGS galaxy NGC~337.
The dashed-blue line shows the unattenuated stellar spectrum 
and the black line shows
the total SED (i.e. stellar emission attenuated by dust + infrared dust emission).
In the bottom panel, we plot the fit residuals,
$(L_\mathrm{obs}-L_\mathrm{mod})/L_\mathrm{obs}$, 
where $L_\mathrm{obs}$ is the observed luminosity and $L_\mathrm{mod}$.}
\label{fig1}
\end{center}
\end{figure}

\begin{figure}
\begin{center}
\includegraphics[width=0.9\textwidth]{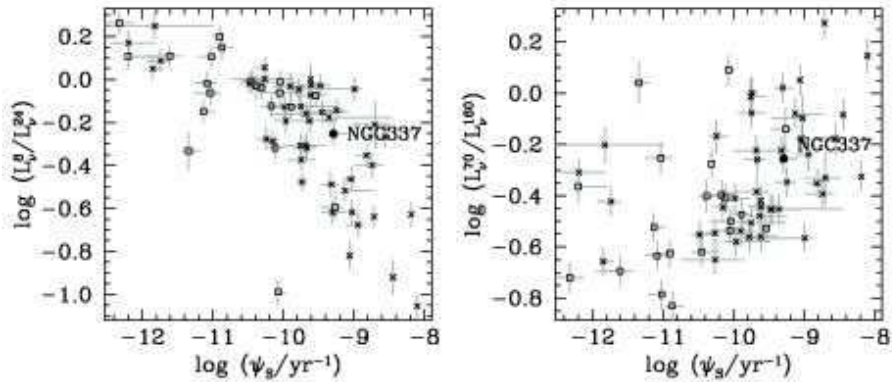}
\vspace*{-0.25 cm} 
\caption{{\it Spitzer} infrared colours ({\it left} -- ratio of 8- to 24-\mic\ luminosity;
{\it right} -- ratio of 70- to 160-\mic\ luminosity) plotted against the median-likelihood estimate of the
specific star formation rate, \ssfr, derived for the galaxies in the SINGS sample.}
\label{fig2}
\end{center}
\end{figure}

We start by exploiting our model to interpret observed SEDs of the SINGS sample,
which contains 75 local galaxies spanning a wide range in morphology and star formation activity. We
compare our model predictions with the ultraviolet, optical and infrared observations of these galaxies
using the method outlined above.
To illustrate the quality of our fits, in Fig.~\ref{fig1}, we show an example of the best-fit SED
of one galaxy of this sample with median {\it Spitzer} infrared colours, NGC~337.

The results of our spectral fits for the full sample show that the mid- and far-infrared colours of galaxies
correlate strongly with the specific star formation rate
(i.e. star formation rate divided by the stellar mass; see Fig.~\ref{fig2}),
as well as with other galaxy-wide properties such as the ratio
of infrared luminosity between stellar birth clouds and the ambient ISM,
the contributions by PAHs and grains in thermal equilibrium to the total
infrared emission, and the ratio of dust mass to stellar mass.

\subsection{SDSS-IRAS sample}

\begin{figure}
\begin{center}
\includegraphics[width=\textwidth]{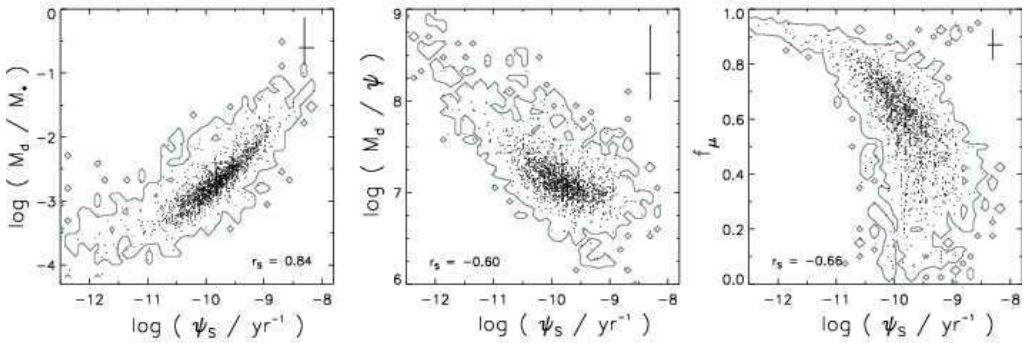}
\vspace*{-0.5 cm} 
 \caption{Median-likelihood estimates of three galaxy
properties against specific star formation rate, \ssfr. {\it Left panel}:
ratio of dust mass to stellar mass, $\mdust/\mstar$. {\it Middle
panel}: ratio of dust mass to star formation rate, $\mdust/\sfr$
(which may be used as a proxy for the dust-to-gas ratio).
{\it Right panel}: fraction of total infrared
luminosity contributed by dust in the ambient ISM, $f_\mu$. In each
panel, the grey contour shows the distribution of the full matched 
\galex-SDSS-2MASS-\iras\ sample described in \cite{daCunha2009},
while the points show the distribution of the sub-sample of 1658 
galaxies with highest-S/N photometry in all bands.
The error bars represent the median confidence
ranges in each parameter. The Spearman rank coefficient $r_{S}$
is indicated in each panel.}
\label{fig3}
\end{center}
\end{figure}

We further investigate the relations between star formation activity and dust content
using a sample of 3258 star-forming SDSS galaxies with complementary
photometric observations by \galex, 2MASS and \iras\ (da Cunha et~al. 2009).
We find that the specific star formation rate correlates strongly with the dust-to-stellar mass
ratio, the ratio of dust mass to star formation rate and the fraction of dust luminosity
contributed by the diffuse ISM (Fig.~\ref{fig3}). A comparison with recent models of chemical
and dust evolution of galaxies suggests that these correlations could arise, at least in part,
from an evolutionary sequence (da Cunha et~al. 2009).

\begin{acknowledgments}

E. da Cunha acknowledges the IAU for a travel grant and also financial support
from the EU ToK grant 39965 and FP7-REGPOT 206469. 

\end{acknowledgments}





\end{document}